\documentclass[journal]{IEEEtran}

\ifCLASSINFOpdf
  \usepackage[pdftex]{graphicx}
\else
  \usepackage[dvips]{graphicx}
\fi
\usepackage[cmex10]{amsmath}
\usepackage{algorithmic}

\usepackage{array}

\ifCLASSOPTIONcompsoc
\usepackage[caption=false,font=normalsize,labelfont=sf,textfont=sf]{subfig}
\else
\usepackage[caption=false,font=footnotesize]{subfig}
\fi
\usepackage{fixltx2e}
\usepackage{amsfonts}
\usepackage{amssymb}
\usepackage{amsthm}
\usepackage{bm}
\usepackage[ruled]{algorithm2e}
\usepackage[utf8]{inputenc}
\usepackage{booktabs}
\usepackage{url}
\usepackage{cite}
\usepackage{xcolor}

\newcounter{tempEquationCounter}
\newcounter{thisEquationNumber}

\renewenvironment{IEEEbiography}[1]
  {\IEEEbiographynophoto{#1}}
  {\endIEEEbiographynophoto}

\begin{document}
%
\title{Realizing the Tactile Internet: Haptic \\Communications over Next  Generation \\5G Cellular Networks}

%
%
%

\author{Adnan~Aijaz,~\IEEEmembership{Member,~IEEE,}
        Mischa~Dohler,~\IEEEmembership{Fellow,~IEEE,}
        A.~Hamid~Aghvami,~\IEEEmembership{Fellow,~IEEE,}
        Vasilis~Friderikos,~\IEEEmembership{Member,~IEEE,}
        and~Magnus~Frodigh
\thanks{
A. Aijaz is now with the Telecommunications Research Laboratories, Toshiba Research Europe Ltd., Bristol, UK. This work was done while he was working at the Centre for Telecommunications Research, King's College London, London, UK. Contact email: adnan.aijaz@toshiba-trel.com

M. Dohler, A. H. Aghvami, and V. Friderikos are with the Centre for Telecommunications Research, King's College London, London, UK. 

M. Frodigh is with Ericsson Research, Stockholm, Sweden.

}}

%
%

\markboth{IEEE WIRELESS COMMUNICATIONS - Accepted for Publication}%
{Shell \MakeLowercase{\textit{et al.}}: Bare Demo of IEEEtran.cls for Journals}
%



\maketitle

\begin{abstract}
Prior Internet designs encompassed the fixed, mobile and lately the ``things'' Internet. In a natural evolution to these, the notion of the \emph{Tactile Internet} is emerging which allows one to transmit touch and actuation in real-time. With voice and data communications driving the designs of the current Internets, the Tactile Internet will enable \emph{haptic communications}, which in turn will be a paradigm shift in how skills and labor are digitally delivered globally. Design efforts for both the Tactile Internet and the underlying haptic communications  are in its infancy. The aim of this article is thus to review some of the most stringent design challenges, as well as proposing first avenues for specific solutions to enable the Tactile Internet revolution.

\end{abstract}


%
\IEEEpeerreviewmaketitle

\section{Introduction}
%
%
%
%
\IEEEPARstart{E}{ach} Internet generation was believed to be the last, with designs pushed to near perfection. The first and original Internet, a virtually infinite network of computers, was a paradigm changer and went on to define the economies of the late 20th century.  However, after that Internet came the \emph{Mobile Internet}, connecting billions of smart phones and laptops, and yet again redefining entire segments of the economy in the first decade of the 21st century. Today, we witness the emergence of the \emph{Internet of Things} (IoT), shortly to connect trillions of objects and starting to redefine yet again various economies of this decade.

These different embodiments of the Internet will be dwarfed by the emergence of the \emph{Tactile Internet}\footnote{The term Tactile Internet had recently been coined by Prof. Gerhard Fettweis  \emph{et al.} \cite{TI}. \textcolor{black}{Whilst the term Haptic Internet would have been a more rigorous term in this context, we shall use the term accepted by the community.}}, in which ultra-responsive and ultra-reliable network connectivity will enable it to deliver physical   \textcolor{black}{ haptic} experiences remotely. \textcolor{black}{The Tactile Internet will add a new dimension to human-machine interaction through building real-time interactive systems.}


Currently, the traditional wired Internet and the Mobile Internet are widely used for delivering content services such as voice telephony, text messaging, video streams, file sharing, emails, etc. The transition towards the IoT is creating a new paradigm of ``control'' communications. However, the Tactile Internet provides a true paradigm shift from content-delivery to skillset/labor-delivery networks, and will thereby revolutionize almost every segment of the society. As discussed in \cite{TI}, the Tactile Internet will enable  remote monitoring and surgery, wireless controlled exoskeletons, remote education and training, remote driving, industrial remote servicing and decommissioning, synchronization of suppliers in smart grid -- among many of its application areas.

Because the Tactile Internet will be servicing really critical aspects of society, it will need to be ultra-reliable and have sufficient capacity to allow large numbers of devices to communicate with each other simultaneously. It will also need to support very low end-to-end latencies as otherwise the tactile user will experience of "cyber-sickness", something observed with gamers and people using flight simulators over poor networks. The Tactile Internet will be able to interconnect with the traditional wired internet, the mobile internet and the internet of things – thereby forming an Internet of entirely new dimensions and capabilities.

At the very core of the design of the Tactile Internet is the 1ms-Challenge, i.e. achieving a round-trip latency of 1 ms at an outage of about 1 ms per day. Realizing the 1ms-challenge would enable the typical latencies and reliabilities required for real-time haptic interaction underpinning unrivalled mobile applications capable of steering and controlling real and virtual objects. Given state-of-the-art 4G mobile/cellular networks have a latency in the order of $20$ ms, a key requirement for fifth generation (5G) mobile networks is to support a round-trip latency of  $1$ ms \cite{5G-GF,5G}, i.e. an order of magnitude faster than 4G.

\textcolor{black}{The conventional Internet facilitates voice and data communications, and provides the medium for audio/visual transport. However, the Tactile Internet will enable \emph{haptic communications} \cite{HC} as the primary application and provide the medium for transporting touch and actuation in real-time i.e., the ability of haptic control through the Internet, in addition to non-haptic control and data (like video and audio). Typically, haptic information is composed of two distinct types of feedbacks: \emph{kinesthetic} feedback (providing information of force, torque, position, velocity, etc.) and \emph{tactile}\footnote{\textcolor{black}{The tactile feedback should not be confused with the Tactile Internet.}} feedback (providing information of surface texture, friction, etc.). The former is perceived by the muscles, joints, and tendons of the body whereas the latter is consumed by the 	mechanoreceptors of the human skin. While the exchange of kinesthetic information closes a global control loop with stringent latency constraints, this is typically not the case with the delivery of tactile impressions. In case of non-haptic control, the feedback is only audio/visual and there is no notion of a closed control loop.  In addition to enabling haptic/non-haptic control/data, the Tactile Internet will enable \emph{networked control systems} (NCS), wherein sensors and actuators are connected and highly dynamic processes are controlled. The control and feedback signals are exchanged in the form of information packets through the network, closing a global control loop and leading to strict latency constraints. Our focus in this paper however is strictly on haptic control as it is inherent to the majority of envisioned Tactile Internet applications.  }

\textcolor{black}{Against this background, the objective of this article is to identify the cutting-edge challenges in realizing the Tactile Internet from both haptic and networking perspectives}. To this end, we begin our discussion with an overview of the potentials and requirements of haptic communications in context of the Tactile Internet architecture. We then translate these into Tactile Internet design challenges with specific emphasis on next generation cellular networks. We also provide practical recommendations for successfully addressing some of the highlighted challenges.

\begin{figure*}
\centering
\includegraphics[scale=0.34]{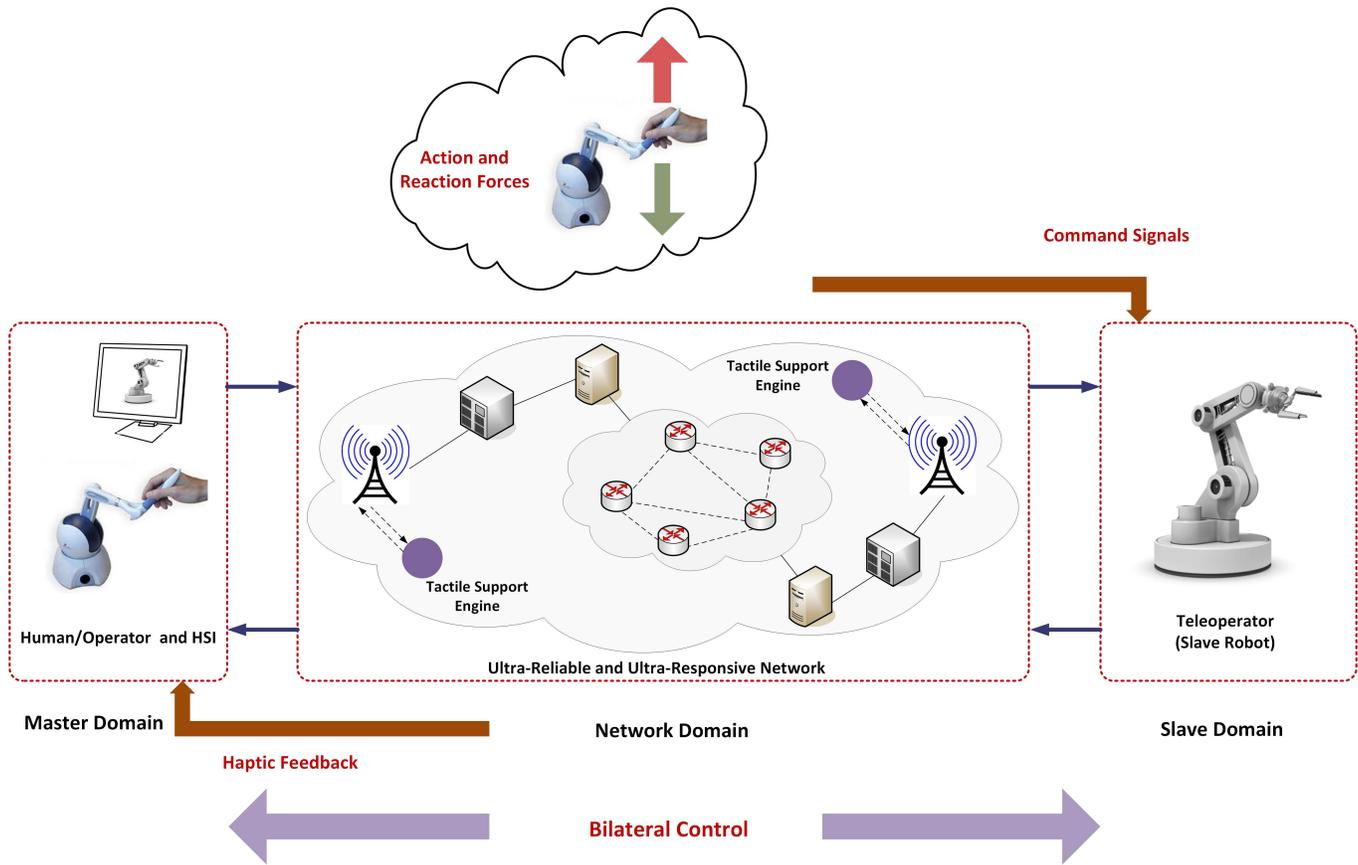}
\caption{ Functional architecture of the Tactile Internet \textcolor{black}{providing the medium for haptic transport}. }
\label{haptic_comm}
\end{figure*}

\section{\textcolor{black}{Towards a Tactile Internet Architecture}}
The haptic sense (sense of touch) establishes a link between humans and unknown environments in a similar way as the auditory and visual senses. Differing from these senses, the haptic sense occurs bilaterally i.e., a touch is sensed by imposing a motion on an environment and feeling the environment by a distortion or reaction force. Haptic communications provides an additional dimension over traditional audiovisual communication for truly immersive steering and control in remote environments \cite{HC}.

\textcolor{black}{As shown in Fig. \ref{haptic_comm}, the end-to-end architecture for the Tactile Internet can be split into three distinct domains: the master domain, the network domain, and the slave domain. The master domain usually consists of a human (operator) and a human system interface (HSI). The HSI is actually a haptic device (master robot), which converts the human input
to haptic input through various  coding techniques. The haptic device allows a user to touch, feel, and manipulate objects in real and virtual environments, and primarily controls the operation of the slave domain.}

\textcolor{black}{The network domain provides the medium for bilateral communication between master and slave domains, and therefore \emph{kinesthetically} couples the human to the remote environment. Ideally, the operator is completely immersed into the remote environment.}

\textcolor{black}{The slave domain consists of a teleoperator (slave robot) and is directly controlled by the master domain through various command signals. The teleoperator interacts with various objects in the remote environment. Typically, no \emph{a priori} knowledge exists about the environment. Through command and feedback signals, energy is exchanged between the master and slave domains thereby closing a global control loop.}

\textcolor{black}{The tactile support engines located closer the edge of the network, as shown in Fig.  \ref{haptic_comm}, provide artificial intelligence capabilities that play a critical role in stabilizing the overall system as discussed later. }

\textcolor{black}{It is important to distinguish between the Tactile Internet and haptic communications. Similar to traditional multimedia (voice, data, video, etc.) communications running over the wired and mobile Internets, the primary application running over the Tactile Internet would be haptic communications. Therefore, haptic communications and the Tactile Internet have a service and medium relationship, much along the lines, for example, of the relation between VoIP and the Internet.}


\section{Research Challenges for the Tactile Internet}
In this section we outline the key research challenges and open problems in realizing the Tactile Internet from both haptic as well as networking perspectives.

\subsection{Haptic Devices}
Haptics is enabled by haptic devices which allow a user to touch, feel, and manipulate objects in real or virtual environments. Such haptic devices are now commercially available e.g., vendors like Geomagic and Sensable have introduced devices with up-to 6 degrees of freedom (DoF). The most popular design for haptic devices is a linkage-based  system which consists of a robotic arm attached to a stylus. The robotic arm tracks the position of the stylus and is capable of exerting a force on its tip. To truly realize the vision of the Tactile Internet, further developments on haptic devices are needed; particularly in increasing the DoF to meet the demands of envisioned applications and embedding the network interface for direct or indirect communication with the cellular network. Besides, the cost of such devices must be reduced for widespread adoption. \textcolor{black}{Finally, most devices offer kinesthetic control only, and therefore, more research is needed to offer both kinesthetic and tactile feedbacks on same device.}

\subsection{\textcolor{black}{Haptic Codecs}}
\textcolor{black}{Over the last decade, numerous studies have appeared in literature on transmitting haptic information, mainly in telepresence systems. As part of digitizing the haptic information, the haptic signals are typically sampled at 1 kHz leading to a fairly high packet generation rate of 1000 packets per second. Considering typical operation in bandwidth-limited networks, different techniques have been investigated for haptic data compression by exploiting the limits of human haptic perception.  However, further developments are needed for realizing the vision of the Tactile Internet.}

\textcolor{black}{A fundamental challenge in context of the Tactile Internet is the development of a \emph{standard} haptic codecs family, similar to the state-of-the-art audio (ITU-T H.264) and video (ISO/IEC MPEG-4) codecs. Embracing both kinesthetic as well as tactile information, such a codec family would be a key enabler for scalability at the network edge and universal uptake. Besides, it introduces a layered approach to haptic data (comprising multi-modal sensory information), which would be crucial for operation in typically challenging wireless environments.}

\subsection{\textcolor{black}{Multi-Modal Sensory Information}}
\textcolor{black}{Besides the haptic feedback, the Tactile Internet must account for provisioning of audio and visual feedbacks, primarily at the master domain. This is because the human brain integrates different sensory modalities \cite{vh-integ} that leads to increased perceptual performance. A key challenge in this context is the cross-modal asynchrony which arises due to the fact that different modalities (visual, auditory, and haptic) have different requirement in terms of sampling, transmission rate, latency, etc. Therefore, a multiplexing scheme is required which is capable of exploiting priorities as well as temporal integration of different modalities. Some initial research efforts aim at addressing this challenge. For example, in \cite{admux}, the authors propose an adaptive multiplexer, termed as Admux, which integrates different modalities in a statistically optimal manner. However, performance of Admux under dynamically changing environments such as wireless, and in particular the error-resiliency to packet losses, has not been investigated. Similarly, in \cite{vh-mux} the authors propose a visual-haptic multiplexing scheme. However, the proposed scheme works specifically over constant bitrate channels. Therefore, further developments in this area are needed for truly immersive steering and control envisioned for the Tactile Internet.}

%

\subsection{\textcolor{black}{Stability for Haptic Control}}
In a haptic communication system, energy is exchanged between the HSI and the teleoperator through command and feedback signals, thereby, closing a global control loop involving the human, the communication (cellular) network, and the remote environment. Hence, stability comes as a natural challenge for the development of control design. It is a well-known fact that communication induced artifacts lead to instability of a control loop system. The issue of instability become important in wireless environments where \emph{time-varying} delays and packet losses are dominant. Instability of a haptic communication system strongly deteriorates the immersiveness into the remote environment. In recent years, several control architectures have been proposed to stabilize haptic systems under \emph{constant} time delays. However, the issue of time-varying delays remains widely unaddressed. For realizing the Tactile Internet, stability of haptic communication systems becomes a key challenge that requires sophisticated approaches beyond traditional control methods as well as joint design of communication protocols and control architectures to account for these issues.


\subsection{\textcolor{black}{Ultra-Reliability}}
Reliability refers to availability/provisioning of certain level of communication service nearly $100\%$ of the time. In cellular networks, reliability is impaired due to a number of factors \cite{URC} such as uncontrollable interference, decreased power of the useful signal, resource depletion, equipment failure, etc.

The Tactile Internet is expected to service key areas of the society, and therefore requires ultra-reliable network connectivity. The term ultra-reliable can be quantified in terms of fixed-line carrier-grade reliability of seven nines i.e., an outage probability of $10^{-7}$, which translates to milliseconds of outage per day. Ultra-reliable network connectivity is critical in keeping packet losses to a minimum. In lossy environments, haptic communications is vulnerable to different types of artifacts resulting in undesirable strong forces and surface roughness (erroneous sensation of being in contact with a significantly rough surface). Such artifacts not only impair the transparency of the system but also directly interfere with operator's activity.

In \cite{URC}, an \emph{Ultra-Reliable Communication} (URC) mode is proposed for 5G cellular networks. The URC mode is built around the concept of \emph{Reliable Service Composition} (RSC) which refers to graceful degradation of service quality in worse communication conditions rather than absolute availability/unavailability. This implies that the communication network can offer a certain level of functionality for the service even when it is not possible to achieve full-functionality. While conventional voice and video applications naturally allow for such graceful degradation (e.g., scalable video coding), it may not work for haptic applications. This is due to the fact that delayed arrival or loss of critical haptic data may lead to instability of the system. Therefore, carrier-grade reliability for the transport of haptic sessions over cellular networks becomes a key challenge in realizing the Tactile Internet.

Such stringent reliability requirements, in turn, require a revisit of the conventional protocol stack due to specific requirements of haptic communications. From a medium access control (MAC, Layer 2) per-link point of view, reliability has to be provided through mechanisms different from ARQ and H-ARQ due to the extra delay it would incur if provided in time. From an end-to-end point of view, the use of the Transmission Control Protocol (TCP) would be desirable at the transport layer.  However, TCP provides high reliability at expense of high protocol overhead. Moreover, the packet rate in haptic communications is generally not flexible, hence the congestion control of TCP is not appropriate, which also results in higher latency. On the other hand, User Datagram Protocol (UDP) provides a low-overhead alternative at the expense of reduced reliability.

Apart from the transmission of actual haptic transport streams, exchange of session information is essential. Unlike the haptic transport stream, the transmission of session information is not constrained by hard delay requirements. Compared to audio/visual session establishment, haptic sessions involve a large number of parameters to be exchanged. However, reliable exchange of session parameters becomes particularly important for properly configuring the haptic system.

Another important issue from the protocol stack perspective is the small packet header to payload ratio. Haptic data is typically sampled at a constant rate with a resolution of $16$ bit per DoF. Therefore, the payload of one packet for a 3-DoF haptic stream is only $6$ bytes. With the ongoing transition towards IPv6, the issue becomes particularly challenging as the header size doubles from $20$ bytes (in IPv4) to $40$ bytes. In literature, a number of header compression techniques have been proposed, both for the wired Internet (IETF RFC 3096)  as well as for the wireless links (IETF RFC 3545). Such techniques provide significant compression of the header from $40$ bytes to less than $2$ bytes.  However, they have been designed for  specific Internet protocols such as RTP, and especially TCP, and might suffer significantly from unreliable wireless links. Besides, there is an inherent trade-off between header compression and the inherent delay this process might entail. Hence, for network-based haptic communications, this issue needs to be revisited.



To summarize, haptic transport requires a reliable protocol stack with minimal protocol overhead and optimized with respect to the specific requirements of haptic communications.

\subsection{\textcolor{black}{Ultra-Responsive Connectivity}}
The Tactile Internet requires a round-trip latency of $1$ ms, which is a mammoth task, and itself needs a number of challenges to be addressed. From the Physical layer perspective, each packet must not exceed a duration of $33$ $\mu$s \cite{5G-GF} in order to enable a one-way Physical layer transmission of $100$ $\mu$s. However, the modulation used in LTE cellular networks is not viable to achieve this requirement as each OFDM symbol is approximately $70$ $\mu$s long. A shorter Transmission Time Interval (TTI) is also desirable to reduce over-the-air latency. However, shorter TTI requires higher available bandwidth. Therefore, the Physical layer in 5G must be designed to cater for such critical requirements.

Each contributing factor in the end-to-end latency must be optimized to achieve the target latency requirements of the Tactile Internet. The air-interface latency is dominated by the fixed control-plane and user-plane latencies. To reduce these latencies, optimizations at different layers of the protocol stack below IP layer are required. The backhaul and core network latency is primarily operator dependent i.e., the choice of the transport network. On the other hand, core Internet latency is variable and largely dictated by queueing delays and routing policies. To summarize, innovations in the air interface, protocol stack, hardware, backhaul, core Internet, as well as in the overall network architecture are needed to meet this challenge.

\textcolor{black}{Whilst the advances on hardware, protocols, and architecture are paramount in diminishing end-to-end delays, the ultimate limit is set by the finite speed of light which sets an upper bound on the maximum separation between the tactile ends.
}

%
%

\subsection{Radio Resource Allocation}

Radio resource management is a key feature of cellular networks. Radio resource allocation, which is a key component of radio resource management, has a direct impact on throughput, latency, reliability, quality-of-service (QoS), and the performance of higher layers.  With the introduction of haptic communications into cellular networks, radio resource allocation becomes particularly challenging as available resources are shared between haptic applications and other human-to-human (H2H) or machine type communications (MTC) applications, having different and often conflicting service requirements.

\textcolor{black} {Due to stringent latency requirements, radio resources must be provided on priority for haptic communications. To provide high tracking performance between master and slave domains, joint resource allocation in the uplink (UL) and the downlink (DL) is necessary. Besides, haptic communications requires symmetric resource allocation with minimum constant rate guaranteed in the UL and the DL owing to its bidirectional nature.}

\textcolor{black}{Novel resource allocation approaches are needed to cater for the requirements of haptic communications. Besides, for the co-existence of haptic and other vertical applications, flexible approaches to radio resource management, capable of providing \emph{on-demand} functionality, would be needed in 5G networks. }

%
%

\subsection{Collaborative Multi-User Haptic Communications}
In collaborative multi-user haptic communications (CMuHC), multiple users interact in a shared remote environment. CMuHC will enable unprecedented powerful applications and revolutionize the way we interact in the cyberspace.

From a networking perspective, CMuHC will inevitably require the formation of a peer-to-peer overlay in order to orchestrate the participation of multiple users in addition to the facilitation of a number of other necessary functions pertaining to the tasks of overlay maintenance and operation \cite{overlay}.  Such overlay creation generates a number of additional challenges on low latency haptic communications since overlay routing and IP-level routing might not be congruent entailing further delays. Also, the degree of decentralization and co-ordination between peers, i.e., the architecture of the peer-to-peer network, will play an important role in the performance of the multi-user haptic application as well as the (routing) distance between peers.

\subsection{\textcolor{black}{Area-Based Sensing and Actuation}}

\textcolor{black}{State-of-the-art haptic devices mostly provide single-point end effectors i.e., single contact point for kinesthetic and tactile feedback. However, human beings perceive touch-based sensations across surfaces, such as the palm of the hand or parts of the body. To achieve the next level of immersion, a distributed or area-based sensing and actuation is required on the haptic device end. Although initial prototypes for haptic devices with multiple contact points (e.g., the \emph{CyberGrasp System}) as well as for deformable artificial skin with built-in distributed sensor systems exist (e.g., \cite{softskin}), further developments would be crucial in realizing the vision of the Tactile Internet. This, in turn, will have an important impact onto communications requirements due to increased rates and a different perception in case of data loss.}

\subsection{\textcolor{black}{Objective Quality Metrics}}
\textcolor{black}{Quality-of-Experience (QoE) evaluation for haptic communications is currently carried out through \emph{subjective} tests  which are performed with the involvement of human testers. Subjective testing is usually expensive to conduct. Besides, achieving high credibility is difficult owing to dependence on various factors including appropriate selection of testers, sample size, environment, etc. On the other hand, evaluation of QoE through \emph{objective} testing is based on the measurement of several parameters related to service delivery. Objective QoE evaluation for haptic communications is widely unaddressed in literature. However, to accurately capture the QoE through objective testing, further developments are needed such as mapping of network performance metrics (intrinsic QoS parameters) to user experience related parameters (e.g., haptic perception) and incorporation of sophisticated models for human haptic control into the objective quality metrics, and development of joint metrics for auditory, visual, and haptic modalities.}

\begin{figure}
\centering
\includegraphics[scale=0.45]{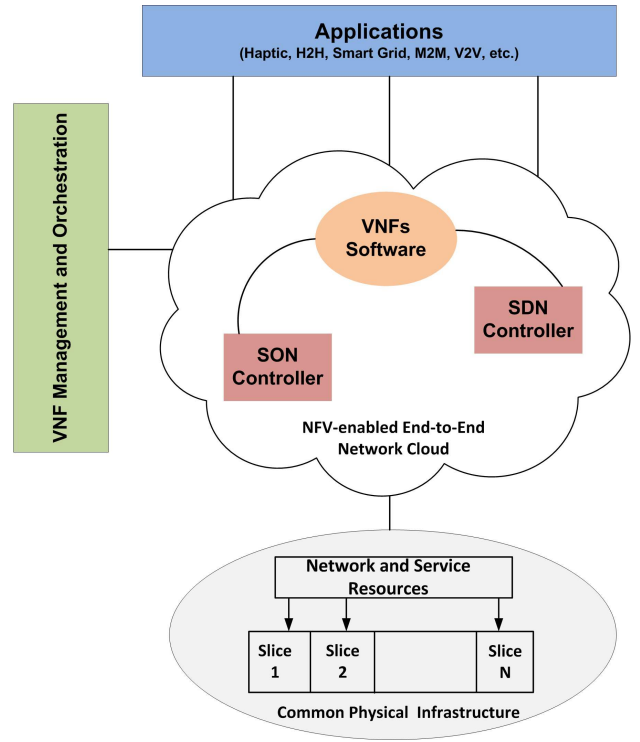}
\caption{ A logical \textcolor{black}{approach to 5G }network architecture based on a common physical infrastructure.}
\label{5G_arch}
\end{figure}

\section{Addressing the Tactile Internet Challenges}
In this section, we provide recommendations for addressing some of the challenges related to the Tactile Internet design highlighted above.

\subsection{Enabling Architecture \textcolor{black}{for Network Slicing}}

In the light of above discussion, one might be persuaded to consider the idea of having a separate network, specifically designed for haptic communications. However, this is not feasible considering capital (CAPEX) as well as operational (OPEX) expenditures. \textcolor{black}{The industry has a general consensus that 5G networks must be designed in a flexible manner such that one network, based on a common physical infrastructure, is efficiently shared among different vertical applications (such as haptic, smart grid, machine-to-machine (M2M), vehicular-to-vehicular (V2V), etc.) to meet the diverse requirements of different applications. Such sharing will be possible through greater degree of abstraction of 5G networks wherein different network \emph{slices} would be allocated to different vertical application sectors. A network slice is defined as a connectivity service based on various customizable software-defined functions that govern geographical coverage area, availability, robustness, capacity, and security \cite{eric_wp}. Such slicing approach provides more of a \emph{network on demand} functionality.}


In realizing this type of network architecture based on a common physical infrastructure, two technologies would be of critical importance: network function virtualization (NFV) and software defined networking (SDN). Both technologies\footnote{Detailed surveys on NFV and SDN are beyond the scope of this paper. Interested readers are referred to \cite{NFV,SDN} and the references therein. } provide the tools to design networks with greater degree of abstraction, increasing the network flexibility.  NFV provides the separation of network functions from the hardware infrastructure. The network function can be managed as a software module that can be deployed in any standard cloud computing infrastructure. On the other hand, SDN provides an architectural framework wherein control and data planes are decoupled, and enables direct programmability of network control through software-based controllers.   Although SDN is viewed as a tool for 5G core network, it can be extended to the radio access part as well in the form of self-organizing networking (SON) solutions \cite{SDN_cellular}.

By unifying NFV, SDN, and SON, we propose a novel \emph{logical} network architecture\footnote{Our focus here is strictly on logical network architecture for 5G. For other aspects of 5G including use cases, scenarios, technology components, spectrum issues, standardization, etc. we refer the interested readers to the flagship EU METIS project (\cite{METIS} and the references therein). } for 5G, which is shown in Fig. \ref{5G_arch}. The proposed architecture is built on a common programmable physical infrastructure and an NFV-enabled end-to-end network cloud that provides all protocol stack functionalities.  The software implementation of network functions, termed as Virtualized Network Functions (VNFs) software, is deployed on the underlying infrastructure. The VNF management and orchestration framework is used to monitor, manage, and troubleshoot VNFs software. The SDN and SON controllers provide the functionality of programming the core and radio access network, respectively. The combination of SDN, SON, and NFV enables \textcolor{black} {flexible and dynamic} slicing of end-to-end network and service resources, which is particularly attractive to cater for the requirements of different vertical applications, including haptic communications, in a flexible manner.

%

\subsection{Reducing End-to-End Latency}
Without any doubt, OFDM is the primary Physical layer candidate for 5G. Over the last few years, several variants of OFDM have appeared in literature that aim to address its potential shortcomings. In \cite{5G}, the authors highlight the concept of \emph{tunable} OFDM for 5G. The key benefit of tunable OFDM is its adaptability for meeting different 5G requirements. In channels with small delay spreads (e.g., millimeter wave channels), the subcarrier spacing could increase and the FFT block size and cyclic prefix can be significantly reduced to achieve lower latency for the Physical layer transmission.

In order to reduce air-interface latency, both the control and the data planes need to be optimized. With reference to the control plane, an important issue on the air-interface is of radio link failures which may occur due to a number of factors and frequently result in loss of RRC connection. To ensure the stability of haptic system, the eNodeB must support a \emph{fast} RRC connection re-establishment feature. This can be achieved by optimizing the random access procedure e.g., contention free access with some dedicated resources, and by optimizing the the RRC connection re-establishment phase through reducing the number of control messages exchanged with the eNodeB. Alternatively, for haptic sessions, the RRC state should be transparent to radio link failures and always stay in the the connected mode after initial session establishment.

In the user-plane, HARQ is used to provide link-level reliability. However, HARQ is not suitable for haptic communications owing to its increased retransmission delay. By disabling HARQ for haptic communications, for reduced air-interface latency, link-level reliability must be provided through some other techniques.

One way of reducing the backhaul delay is to adopt optical transport as the backhaul medium. An attractive alternative to deploying optical fiber is a full-duplex wireless backhaul, especially in higher spectrum bands. Due to the distinct characteristics of full-duplex communications, full-duplex wireless backhaul can be realized in two distinct ways: (a) bi-directional link between the eNodeB and the core network, and (b) two unidirectional links; one from the user to the eNodeB and the other from the eNodeB to the core network.

Last, but not the least, it is important to reduce the processing delay in different nodes of the networks. The processing delay can be reducing by increasing the computational power of different nodes/entities. In \cite{TI}, Highly Adaptive Energy Efficient Computing (HAEC) boxes have been envisioned. HAEC boxes provide an exa-scale of computing power and therefore, would play an important  role in reducing processing delays in 5G networks.

\subsection{Achieving Ultra-Reliable Connectivity}
Ultra-reliable connectivity is not just important for haptic data transfer but also for exchange of session parameters for proper system configuration. Improving wireless connectivity to carrier-grade standard is not an invincible task. By exploiting frequency diversity over multiple uncorrelated links, carrier-grade reliability can be approached. Multiple uncorrelated links can be  created through inter-band spectrum aggregation techniques as well as through coordinated multi-point (CoMP) \cite{CoMP} transmissions.


Due to stringent latency requirements of haptic communications, diversity in the time domain is not feasible and therefore, techniques like HARQ cannot be applied. A possible solution for link-level reliability while eliminating re-transmissions is to provide HARQ-like functionality in the frequency or spatial domain by exploiting the concept of multiple uncorrelated channels.

\begin{figure}
\centering
\includegraphics[scale=0.38]{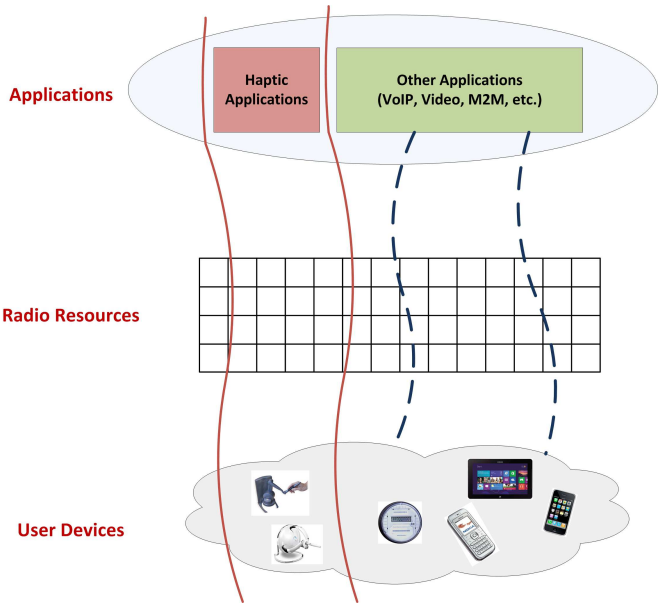}
\caption{An illustration of the slicing approach based on virtualization of radio resources.}
\label{vra}
\end{figure}

\subsection{\textcolor{black}{Virtualization-based Radio Resource Allocation Scheme}}
With respect to the challenges of low latency, high reliability, and radio resource allocation, we propose a flexible resource management scheme, which is based on the proposed logical architecture and the virtualization of radio resources. Virtualization enables flexible slicing, isolation\footnote{Isolation means that any change in one slice due to channel conditions, user mobility, etc. should not result in reduction of radio resources across other slices}, and customization of radio resource across different applications and user devices. \textcolor{black}{Such an approach not only ensures resources are allocated to tactile applications on priority but also facilities the implementation of tactile-specific scheduling algorithms. The key steps of the proposed scheme are as follows.}

\textcolor{black}{\emph{Slicing} -- Initially radio resources are allocated to different slices. Generally either \emph{bandwidth-based} or \emph{resource-based} provisioning is used for allocating radio resources to different slices. In bandwidth-based approach, resource allocation to each slice is defined in terms of aggregate throughput that will be obtained by the flows of that slice. On the other hand, in resource-based approach, a fraction of base station's resources are allocated to each slice. We suggest a combination of bandwidth-based and resource-based provisioning of resources to different slices that results in efficient utilization of resources.}

\textcolor{black}{\emph{Isolation} -- We propose an end-to-end isolation of resources between haptic applications/users and other applications/users as shown in Fig. \ref{vra}. For other applications/users, resource can be managed in two distinct ways: (a) isolation of resources across user devices but not across applications, and  (b) isolation of resources across applications but not across user devices. Such isolation scheme not only guarantees availability of resources for haptic applications, but also allows for customization of resources for other applications.  }

\textcolor{black}{\emph{Customization} -- In this step, resources are customized according to the service requirements of different applications. For example dynamic scheduling might not be feasible for haptic communications due to the disproportionally large control signalling compared to haptic data. Hence, persistent scheduling schemes can be used wherein resource are allocated to haptic applications/users for a given set of sub-frames.}

\textcolor{black}{\emph{Slicing Period} -- The slicing period is defined as the time after which the size of each allocated slice would be re-calculated.  Slicing period plays an important role in ensuring the maximum utilization of scarce radio resources. In dynamic environments, the size of each slice needs to be adjusted more frequently. Hence, at the beginning of each slicing period the resource slicing process would be repeated. }

%
%

An important challenge in the proposed resource management scheme is the optimal slicing of radio resources between haptic and other applications, which is left as a future work.

\subsection{\textcolor{black}{Edge-Intelligence for Stability of Haptic Control} }
Instability in haptic control occurs primarily due the two key wireless channel impairment: latency and packet loss. Some recent studies show that latency has a higher detrimental effect on stability of the system compared to packet loss.

\textcolor{black}{Recently, there has been a growing trend towards bringing intelligence closer to the edge of cellular networks. Advanced caching techniques and user-oriented traffic management at the edge of the network improves network performance by de-congestion of the core network and reduction of end-to-end latency. Such edge-intelligence techniques will play a critical role in making the Tactile Internet a reality. A practical approach to reduce the impact of latency on haptic control is to deploy predictive and interpolative/extrapolative modules closer to the edge of the network in any advanced cloud infrastructure. Such tactile support engines (illustrated in Fig. \ref{haptic_comm}) enable statistically similar action to be taken autonomously whilst the actual action is on its transmission way via the network. For this purpose, different types of machine learning techniques can be adopted. This approach not only brings stability to the Tactile Internet but also helps in overcoming the fundamental limitation set by the finite speed of light by allowing a wider geographic separation between the tactile ends.}

%

\section{Concluding Remarks}
The Tactile Internet will revolutionize almost every segment of the society. It is a quantum leap prospect for global economy. The Tactile Internet creates daunting new requirements for 5G cellular networks. After introducing the connection between haptic communications and the Tactile Internet, this article  reviewed some of the most important design challenges in realizing the Tactile Internet. The most critical challenge is clearly in providing the 1-ms round-trip latency which we showed to be best accomplished through cutting-edge networking design as well as providing an enhanced haptic perception. The former is best accomplished through fundamental changes to the air interface and architecture design at the wireless edge; whilst the latter is best accomplished through artificial intelligence and predictive analytics able to understand haptic actuation. To enable these more advanced features, a standard haptic codecs family is required which we identified as one of the most interesting research challenges of the coming years.



%



%
%
%
\section*{Acknowledgment}
The authors would like to thank Prof. Gerhard Fettweis, TU Dresden, Germany, for the fruitful discussions on this topic. The healthy criticism, valuable comments, and positive suggestions from anonymous Reviewer 1 are greatly appreciated. 
%
%



\bibliographystyle{IEEEtran}
\bibliography{haptics}
%

%


%
%
\begin{IEEEbiography}{Adnan Aijaz}
(M'14) received the M.Sc degree in Mobile and Personal Communications and the Ph.D degree in Telecommunications Engineering from King’s College London, in 2011 and 2014, respectively. After a post-doc year at King's College London, he moved to Toshiba Research Europe Ltd. where he currently works as a Research Engineer. His research interests include 5G wireless networks, Internet-of-Things, full-duplex communications, cognitive radio networks, smart grids, and molecular communications. He has served as the symposium co-chair for IEEE SmartGridComm'15. He is serving or has served on the TPC of various IEEE conferences and workshops. Prior to joining King's, he worked in cellular industry for nearly 2.5 years in the areas of network performance management, optimization, and quality assurance. He obtained the B.E. degree in Electrical (telecom) Engineering from National University of Sciences and Technology (NUST), Pakistan.
\end{IEEEbiography}
\begin{IEEEbiography}{Mischa Dohler} (S'01–M'03–SM'07-F'14) is full Professor in Wireless Communications at King's College London, Head of the Centre for Telecommunications Research, co-founder and member of the Board of Directors of the smart city pioneer Worldsensing, Fellow and Distinguished Lecturer of the IEEE, and Editor-in-Chief of the Wiley Transactions on Emerging Telecommunications Technologies and the EAI Transactions on the Internet of Things. He is a frequent keynote, panel and tutorial speaker. He has pioneered several research fields, contributed to numerous wireless broadband, IoT/M2M and cyber security standards, holds a dozen patents, organized and chaired numerous conferences, has more than 200 publications, and authored several books. He acts as policy, technology and entrepreneurship adviser. He has talked at TEDx and had coverage TV \& radio.
\end{IEEEbiography}
\begin{IEEEbiography}{Abdol-Hamid Aghvami}
(M'89--SM'91--F'05) is a Professor of Telecommunications Engineering at King's College London. He has published over 600 technical papers and given invited talks and courses world wide on various aspects of Personal and Mobile Radio Communications. He was visiting Professor at NTT Radio Communication Systems Laboratories in 1990, Senior Research Fellow at BT Laboratories in 1998-1999, and was an Executive Advisor to Wireless Facilities Inc., USA, in 1996-2002. He was a member of the Board of Governors of the IEEE Communications Society in 2001-2003, was a Distinguished Lecturer of the IEEE Communications Society in 2004-2007, and has been member, Chairman, and Vice-Chairman of the technical programme and organising committees of a large number of international conferences.  He is a Fellow of the Royal Academy of Engineering, Fellow of the IET, Fellow of the IEEE, and in 2009 was awarded a Fellowship of the Wireless World Research Forum in recognition of his personal contributions to the wireless world. 
\end{IEEEbiography}
\begin{IEEEbiography}{Vasilis Friderikos} is currently a Senior Lecturer at the Centre for Telecommunications Research, King’s College London. His research interests lie broadly within the closely overlapped areas of wireless networking, mobile computing, and architectural aspects of the Future Internet. He has been organizing committee member of the Green Wireless Communications and Networks Workshop (GreeNet) during VTC Spring 2011. He has been the track co-chair at more than 40 flagship IEEE conferences  over the last 7 years. He has also been teaching advanced mobility management protocols for the Future Internet at the Institut Supérieur de l’Electronique et du Numérique (ISEN) in France during autumn 2010.  He has been visiting researcher at  Rutgers University (USA) and recipient of the British Telecom Fellowship Award in 2005. He has supervised 9 PhD students published more than 130 research papers; he is a member of IEEE, IET and INFORMS section on Telecommunications.
\end{IEEEbiography}
\begin{IEEEbiography}{Magnus Frodigh} was born in Stockholm, Sweden,
in 1964. He received the M.S. degree from Linkping University of Technology, Sweden and the Ph.D. degree in radio communication systems from Royal Institute of Technology, Stockholm, Sweden. Since 2013, he is adjunct Professor at Royal Institute of Technology in Wireless Infrastructures. He is currently Research Area Director for Wireless Access Networks at Ericsson Research, and responsible for research in radio network architecture, protocols and algorithms. The work addresses 5G technologies, LTE, LTE advanced including its continued evolution. He joined Ericsson in 1994 and has since held various key senior positions within Ericsson’s Research and Development and Product Management focusing on 2G, 3G, 4G, and 5G technologies.
\end{IEEEbiography}





\end{document}